\documentclass[aps,prb,superscriptaddress]{revtex4-1} 

\usepackage{amsmath,amssymb}
\usepackage{graphicx}
\usepackage{epsfig}
\usepackage[sort&compress]{natbib}

\begin{document}

\title{Enhanced nonlinear imaging through scattering media using transmission matrix based wavefront shaping}

\author{Hilton B. de Aguiar}
\email{h.aguiar@fresnel.fr}
\affiliation{Aix-Marseille Universit\'{e}, CNRS, Centrale Marseille, Institut Fresnel UMR 7249, 13013 Marseille, France}
\author{Sylvain Gigan}
\affiliation{Laboratoire Kastler Brossel, UMR 8552 of CNRS and Universit\'{e} Pierre et Marie Curie, 24 rue Lhomond, 75005 Paris, France}
\author{Sophie Brasselet}
\email{sophie.brasselet@fresnel.fr}
\affiliation{Aix-Marseille Universit\'{e}, CNRS, Centrale Marseille, Institut Fresnel UMR 7249, 13013 Marseille, France}




\begin{abstract}
Despite the tremendous progresses in wavefront control through or inside complex scattering media, several limitations prevent reaching practical feasibility for nonlinear imaging in biological tissues. While the optimization of nonlinear signals might suffer from low signal to noise conditions  and from possible artifacts at large penetration depths, it has nevertheless been largely used in the multiple scattering regime since it provides a guide star mechanism as well as an intrinsic compensation for spatiotemporal distortions. Here, we demonstrate the benefit of Transmission Matrix (TM) based approaches under broadband illumination conditions, to perform nonlinear imaging. Using ultrashort pulse illumination with spectral bandwidth comparable but still lower than the spectral width of the scattering medium, we show strong nonlinear enhancements of several orders of magnitude, through thicknesses of a few transport mean free paths, which corresponds to millimeters in biological tissues. Linear TM refocusing is moreover compatible with fast scanning nonlinear imaging and potentially with acoustic based methods, which paves the way for nonlinear microscopy deep inside scattering media.
\end{abstract}


\maketitle

\section{Introduction}
Nonlinear microscopy (NLM) is established as a powerful approach for label-free imaging in biological tissues. Despite the impact of NLM on many fields, standard nonlinear imaging modalities can however only image at shallow depths, typically a few hundreds of micrometers for biological specimens. The penetration depth is hindered by optical aberrations and scattering, which degrade the spatial quality of focusing and decrease the spatiotemporal coherence of short pulse excitation fields necessary to build up nonlinear processes. In order to enhance the penetration depth, adaptive optics has been introduced to compensate for aberrations from  biological samples~\cite{Debarre2009}. Although signal enhancements up to two orders of magnitude can be reached, adaptive optics is limited to depths of a few scattering mean free path (typically a millimeter in biological tissues), since it relies on ballistic photons~\cite{Tang2012,Wang2014,Sinefeld2015}. It also most often relies on the optimization of an existing nonlinear signal, which inherently limits this technique in case of strong scattering.

Wavefront shaping (WS) can overcome penetration depths limits in highly scattering media, addressing specifically the scattered photons~\cite{Mosk2012}. Despite the random appearance of the outgoing speckle pattern arising from multiple scattering events, it has nevertheless a deterministic relation with the original incoming wavefront, revealed in the Transmission Matrix (TM) of the scattering medium that connects incoming to outgoing fields. By coherently controlling a speckle pattern, one can increase the energy density at targeted positions, resembling a focus that we call "refocus" in what follows to avoid confusion with a focus obtained from ballistic photons. WS has shown to be able to refocus through~\cite{Vellekoop2007,Popoff2010}, or inside~\cite{Chaigne2013} scattering media, thanks to the manipulation of the wavefront phase or amplitude~\cite{Akbulut2011,Conkey2012}.

The combination of WS and NLM is however so far limited to only a few demonstrations, based on nonlinear feedback optimization~\cite{Aulbach2012,Katz2011,Katz2014b}. These methods offer in particular a way to obtain an intrinsic spatiotemporal control of a refocused spot, by the use of degrees of freedom offered by the intrinsic space-time coupling in scattering media~\cite{Katz2011,Aulbach2012,Katz2014b,Vellekoop2015}. The use of broadband spectral excitation conditions, necessary for nonlinear optical generation, is however limited by the fact that the spectral width of the excitation pulse might be larger than the spectral width of the scattering medium (which is inversely proportional to its Thouless time)~\cite{Paudel2013,Andreoli2015}, leading to a decrease of optimization efficiency due to the need to compensate both spatial dispersion temporal broadening~\cite{Aulbach2012}. Demonstrations using either a nonlinear guide star~\cite{Aulbach2012} or an integrated nonlinear signal~\cite{Katz2014b} have thus led so far to only mild nonlinear enhancements of two orders of magnitude, which is below the optimal expected values.

In this work we address the use of TM-based manipulation to generate nonlinear signals through a scattering medium, under broadband conditions where the spectral width of the excitation is similar or slightly larger than the spectral width of the medium, \emph{i.e.} the spectral correlation width of the speckle. In particular, we take advantage of the fact that spectral decorrelation bandwidths in biological media at 1-2~mm-deep (corresponding to a transport mean free path $l_t$) are comparable with the bandwidth of common laser sources used for NLM (about 100~fs pulse length)~\cite{Katz2014}. We demonstrate that in such regime a TM acquired using either the \emph{linear} or \emph{nonlinear} feedback is compatible with \emph{nonlinear} signal enhancement. We show that linear-feedback-based TM offers fast nonlinear imaging capabilities and high signal-to-noise conditions. This scheme is moreover applicable to nonlinear imaging \emph{inside} a scattering medium, being compatible with acoustics based approaches~\cite{Chaigne2013}. We illustrate the potential of this approach in Second Harmonic Generation (SHG) imaging of thick collagen fibers placed behind scattering media of thicknesses close to $l_t$, in conditions that would fail using nonlinear feedback optimization.

\section{Results}
\begin{figure}[htpb]
    \centering
    \epsfig{file=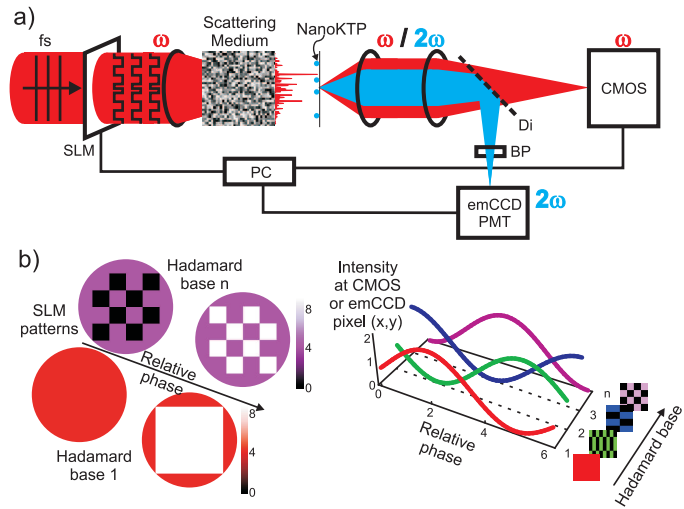,width=0.5\columnwidth}
    \caption{Wavefront shaping experiments for nonlinear microscopy.
		a) Simplified experimental layout. The optical wavefront is shaped by an SLM and focused inside the scattering medium. The speckle generated outside the scattering medium excites the nonlinear sources (nanoKTP) placed at a plane further imaged on different detectors (emCCD, CMOS, PMT).
		b) Methodology used for refocusing using the Hadamard basis. The wavefront phase is swept in respect to a reference field (the unmodified periphery outside a Hadamard base), under broadband illumination condition. At the detectors (a pixel on the emCCD or on the CMOS) a sinusoidal modulation of the intensity is observed, from which each individual phase is stored (per base). After scanning the basis set, the optimal wavefront is used for either enhancing the nonlinear signal (emCCD) or linearly refocusing (CMOS) with the nonlinear signal detected in parallel (PMT).}
    \label{fig1}
\end{figure}

Fig.~\ref{fig1}a illustrates the general experimental layout that we use for both nonlinear and linear feedbacks, both based on TM measurements. In both cases, the excitation is spectrally broadband with a pulse length of about 130~fs (800~nm central wavelength). The incident linear beam wavefront is manipulated by a spatial light modulator (SLM) imaged at the backfocal plane of the focusing lens. A nonlinear SHG-active sample is placed at about 500~$\mu$m distance after the scattering medium, and imaged on either a pixelized sensitive camera or on a large area integrating detector. For acquisition of the TM (Fig.~\ref{fig1}b), we tune the incident wavefront phase, following an approach thoroughly described in Ref.~\cite{Popoff2011a} and detailed in the Methods section. The incident wavefront is decomposed on the Hadamard basis --- instead of the canonical basis --- to benefit from higher fluences at the sample plane. The pixel signal on which the measurement is performed is recorded either at the incident linear wavelength (linear feedback) or at the SHG wavelength emitted from a SHG-active region of the sample (nonlinear feedback). Potassium titanyl phosphate nanocrystals (nanoKTP) of 150~nm diameter are used to monitor the SHG enhancement obtained after both linear and nonlinear feedbacks. These nanocrystals are in particular of superior photostability~\cite{LeXuan2008} as compared to fluorescent samples that are likely to photobleach~\cite{Vellekoop2008c,Katz2014b}. Quantitative analysis is moreover based on single isolated nanocrystals in order to avoid any bias due to multiple interference sources, such as clustered particles.

\subsection{Nonlinear feedback refocusing}
\begin{figure}[htpb]
    \centering
    \epsfig{file=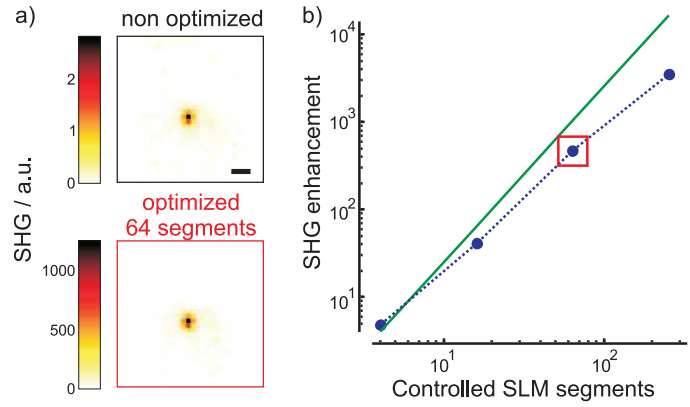,width=0.5\columnwidth}
    \caption{Nonlinear signal enhancement based on a nonlinear feedback on a single 150~nm diameter nanoKTP crystal through a scattering medium (diffuser).
		a) SHG wide field images, taken with the emCCD, before (upper panel) and after WS (lower panel). The non optimized image is an average over random wavefronts, whereas the optimized image is a single frame. Scale bar: $1.6~\mu$m.
		b) Nonlinear SHG enhancement of a single particle versus the number of independently controlled SLM segments (blue markers). The red box refers to the point depicted in a). The green line follows a quadratic dependence of the nonlinear enhancement with respect to linear enhancements estimated after Ref.~\cite{Popoff2010}.}
    \label{fig2}
\end{figure}

Fig.~\ref{fig2}a shows SHG images obtained when a single isolated nanoKTP crystal is placed behind a diffuser, using non-optimized and optimized wavefronts. The optimized image is formed after WS has been performed at the nanoKTP position (see Methods section) using the information obtained from a TM measured using a nonlinear feedback. While linear feedback TM is standard~\cite{Popoff2010}, nonlinear feedback still allows measurement of the phase of the TM elements~\cite{Aulbach2012}. This approach evidences the capacity of the nonlinear feedback-based TM to refocus light on this specific isolated nanocrystal and to generate a high nonlinear response.

Remarkably, up to four orders of magnitude nonlinear enhancements are achieved by using a few hundreds of independently controlled SLM segments (N$_{SLM}$). 
Fig.~\ref{fig2}b shows that the nonlinear enhancement obtained from a single nanoKTP signal depends quadratically on N$_{SLM}$ (the departure at high N$_{SLM}$ is attributed to experimental artifacts such as possible correlations between SLM segments~\cite{Popoff2011a}). This expected dependence ascertains that the refocused spot preserves optimal nonlinear coherent build-up.

The nonlinear enhancements measured here are far above what has been previously reported in multiply scattering media using even higher N$_{SLM}$ values~\cite{Aulbach2012}. This difference is assigned to the regime used here. Even though the excitation pulse is broadband, its spectral bandwidth lies below the spectral bandwidth of the scattering medium, or similarly the pulse length of 130~fs surpasses the Thouless time of the medium. Note that this lengthening time scales as $L^2 / D$, with $D$ the diffusion constant of the medium (proportional to $l_t$), and $L$ the medium thickness~\cite{Paudel2013,Andreoli2015}. These properties can be monitored by the transmitted speckle contrast~\cite{Curry2011,Paudel2013,Andreoli2015}: for the scattering medium used a speckle contrast of 0.98 has been measured, which is indeed the signature of a large medium spectral bandwidth~\cite{Curry2011,Paudel2013}.

Even though nonlinear feedback is able to provide remarkably high enhancements, it still imposes practical constraints. Using 64 controllable SLM segments requires in particular TM acquisition times of the order of minutes, which can be at extreme costs for nonlinear imaging. Furthermore, this methodology also relies on low signals, especially at large depths, and can be only measured in regions where nonlinear signal is emitted. To answer these issues, we address in the next section the use of the overwhelming linearly scattered photons for nonlinear signal enhancements.

\subsection{Linear feedback refocusing}
\begin{figure}[htpb]
    \centering
    \epsfig{file=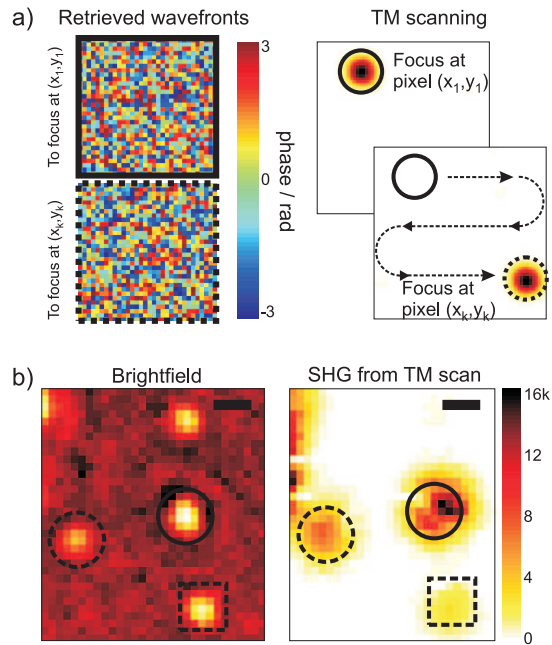,width=0.4\columnwidth}
    \caption{Nonlinear signal enhancement based on a linear feedback through a scattering medium (diffuser).
		a) After acquisition of the linear transmission matrix, wavefronts are tailored to refocus the beam at desired positions through the scattering medium. The optimal wavefronts obtained for two different positions are depicted.
		b) Comparison between a brightfield image (left, revealing the location of the nanoKTP crystals) and the obtained corresponding SHG image (right). The SHG image is obtained by a scanning of the focus positions obtained from the linear feedback optimization procedure. Number of controlled SLM segments: $2^{12}$. Scale bar: $0.9~\mu$m.}
    \label{fig3}
\end{figure}

The principle of SHG image reconstruction by linear feedback is sketched in Fig.~\ref{fig3}. After acquiring the TM of the system, refocus is formed at a desired location, with simultaneous acquisition of nonlinear signals, resembling raster-scanning methodologies. Note that as above, the TM of the medium is measured under spectrally broadband conditions: the measured interferometric modulation is the spectrally integrated modulation measured on the detector. For a medium with negligible temporal broadening, this approach is equivalent to Ref.~\cite{Popoff2010}, whereas for a medium that broadens the pulse, the acquired TM can be seen as the incoherent sum of different spectral contributions. Figure~\ref{fig3}a shows representative incident wavefronts corresponding to the two positions of the refocus, showing no correlation as expected in multiple scattering conditions. The acquisition of the TM procedure for 256 controllable SLM segments takes about 40~s per region of interest (ROI), which is considerably faster than in the nonlinear feedback procedure described above. Note that linear feedback is limited by the SLM speed, rather than signal-to-noise issues as in nonlinear feedback. Alternatively, one could use the memory effect to reconstruct the final SHG image, with a field of view that depends on the scattering medium properties and set-up configuration~\cite{Vellekoop2010,Ghielmetti2014,Katz2014b,Schott2015}.

To validate the imaging mode methodology, we compare in Fig. \ref{fig3}b the SHG image of isolated nanoKTP crystals acquired by refocus scanning, with a brightfield image taken without the scattering medium. The positions of the three detected nanocrystals are in excellent agreement with the brightfield image. The difference in SHG signal levels is due to the relative orientation of the crystal with respect to the polarization state of the excitation field, indicating that the polarization of the refocused beam is also conserved~\cite{Bicout1994,Xu2005,Ghosh2003,deAguiar2015,deAguiar2015a}. Note that the spatial resolution of the final SHG image is ultimately determined by the spatial coherence length of the speckle (the size of the speckle grain). An additional $1/\sqrt{2}$ factor to the resolution comes from the nonlinear nature of the process, as known in NLM conditions. In comparison, the use of a nonlinear feedback might lead to spurious effects if the distribution of nonlinear emitters is complex in the three dimensions of the sample as discussed in Ref.~\cite{Katz2014b}.

\begin{figure}[htpb]
    \centering
    \epsfig{file=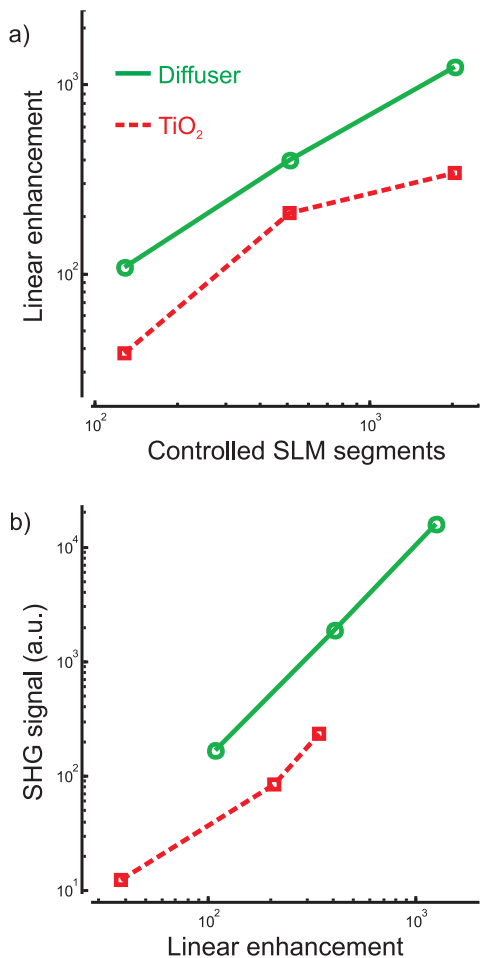,width=0.35\columnwidth}
    \caption{Linear refocusing and nonlinear efficiencies obtained after linear feedback in two different scattering media, diffuser and 20~$\mu$m thick TiO$_2$ film. a) Linear enhancement obtained as a function of the numbers of SLM segments N$_{SLM}$. b) SHG signal dependence on the linear enhancement.}
    \label{fig4}
\end{figure}

Fig. \ref{fig4} summarizes the imaging enhancement performances evaluated on single nanoKTP crystals, for various N$_{SLM}$ values after two scattering media: a surface diffuser and a multiply scattering medium with a thickness of a few $l_t$'s. SHG signals are averaged over several nanoKTPs, circumventing the bias due to orientational effects. While the diffuser visibly shows a large spectral bandwidth as compared to the illumination bandwidth, as mentioned above, the 20 $\mu$m thickness TiO$_2$ film exhibits a speckle contrast of 0.59, which departs from this situation. As a consequence, the linear enhancement values (Fig. \ref{fig4}a) are seen to be higher for the diffuser than the TiO$_2$ film, which is expected from the less favorable number of accessible spectral modes for the later~\cite{Paudel2013}. Under broadband refocusing conditions, the loss in linear enhancement between the two considered media is expected to be about a factor 2 considering the measured speckle contrasts~\cite{Paudel2013}, which is close to what is measured here at relatively low number of segments N$_{SLM}$ (high N$_{SLM}$ values are more sensitive to possible artifacts).

Fig.~\ref{fig4}b shows nonlinear efficiencies generated after refocusing on nanoKTPs in both scattering media. In order to correct for systematic linear enhancements drifts at high N$_{SLM}$ values, we plot the SHG signal versus the linear focus enhancement. Both media exhibit an almost quadratic dependence of the SHG signal with the linear enhancement, evidencing the capabilities of the linear optimization scheme to preserve the nonlinear coupling quality whatever the number of SLM segments used and spatiotemporal coupling conditions. The signals obtained through TiO$_2$ are however an order of magnitude lower than for the diffuser. This is visibly a consequence of its lower spectral bandwidth, which is likely to lead to pulse lengthening and thus to lower nonlinear coupling efficiency. SHG signals are nevertheless of high signal to noise ratio and evidence the remarkable capacity of broadband linear refocusing through scattering media to produce efficient nonlinear conversion.

Importantly in the case of the diffuser, the SHG enhancements are the same as obtained in the nonlinear feedback process described above (SHG enhancements could not be quantified for TiO$_2$ due to the low signal to noise obtained before optimization). This emphasizes that in conditions where medium and laser spectral bandwidth are of similar magnitude, both TM acquisition methodologies lead to efficient spatiotemporal coupling effects, most probably preserving short pulse width after refocusing.

\subsection{Nonlinear bio-imaging behind a scattering medium}
\begin{figure}[htpb]
    \centering
    \epsfig{file=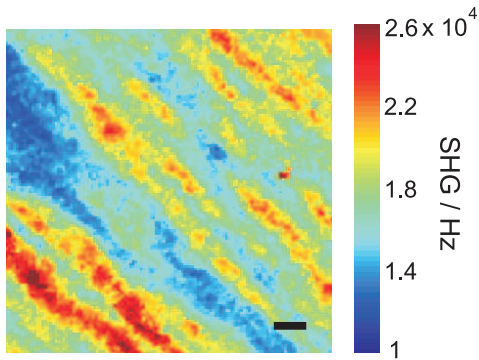,width=0.35\columnwidth}
    \caption{SHG imaging of rat tail tendon collagen. Scattering medium: diffuser. Number of controlled SLM segments: $2^{10}$. Scale bar: $2.3~\mu$m.}
    \label{fig5}
\end{figure}

We finally apply the linear feedback TM measurement to nonlinear imaging in a biological SHG-active thick sample, made of collagen fibers extracted from rat tail tendons (thickness 120 $~\mu$m), a widely studied specimen in SHG microscopy~\cite{Roth1979,Bancelin2014}. Fig. \ref{fig5} shows an SHG image of such sample placed behind a diffuser, acquired after the TM has been measured. To obtain the final SHG image, we first acquired a linear optical image that is used for normalization, allowing correction for possibly remaining spatially non-uniform refocus.

Thick fibers are clearly visible along the diagonal of the image, with a high SHG signal that is clearly above the background signal level. Without WS, a raster-scanning of the focus did not generate a discernible image of the collagen fibers but only a noisy, un-resolved image arising from the speckle excitation. The obtained image is also a remarkable improvement as compared to the use of nonlinear feedback optimization, which gave considerably lower SHG enhancements. This nonlinear sample indeed differ strongly from nanoKTPs since its response is much less sparse: in this regime, the presence of background nonlinear signals from the sample volume and surface is likely to make the nonlinear optimization process biased and much less efficient~\cite{Theer2006,Horton2013}.

\section{Discussion and conclusions}
This study shows that the medium spectral correlation width is a key factor for the manipulation of broadband TM in scattering media, as already evidenced in monochromatic regime in multiple scattering media at a few $l_t$'s~\cite{Muskens2008,Andreoli2015}. Nevertheless in this regime, and under broadband conditions, both linear and nonlinear feedbacks for acquisition of the TM lead to considerable nonlinear enhancements. Beyond such depth in anisotropic scattering media such as in biological tissues (\emph{e.g.} $g \sim$ 0.7-0.99)~\cite{Cheong1990}, the fluence is expected to decrease, thus becoming the limiting factor for nonlinear excitation. However, one could forecast the possibility to generate nonlinear signals using higher energy-per-pulse lasers at reduced repetition rate~\cite{Horton2013}.

Our results show that  refocusing through a scattering medium using the linear TM can allow faster and more efficient nonlinear imaging. Other advantages of the use of linear feedback can be enlightened. First, nonlinear feedback requires fluorescent ~\cite{Katz2011,Katz2014b} or SHG~\cite{Aulbach2012} active guide stars. Under low signal to noise conditions, the feedback mechanism might thus fail or lead to large optimization times. Conversely, the presence of sparse bright regions in the sample can bias the optimization conditions if a large dynamic range of signals is present in the sample~\cite{Katz2014b}. Second, nonlinear feedback is inherent to the nonlinear mechanism used for the optimization procedure. Spatial and temporal properties of the refocus depend on the size of the nonlinear effective volume, on the nonlinear order (two-, three-photon etc.) of the process used, and on its coherence (SHG versus two photon fluorescence for instance; coherent nonlinear effects indeed involve phase matching conditions that might depend on the object size). Optimization for a given nonlinear process is therefore not necessarily appropriate for other nonlinear processes and samples.

In contrast, we anticipate that a linear feedback in hybrid approaches will enable super penetration of NLM. Considerable advances have been made recently towards the direction of linear refocusing \emph{within} scattering media. In particular, acousto-optic~\cite{Xu2011,Wang2012,Judkewitz2013} or photoacoustic~\cite{Kong2011,Chaigne2013,Conkey2015,Lai2015} guide stars have been developed to take advantage of the ballistic penetration of acoustic waves in biological tissues. The use of photoacoustic time traces to monitor the TM inside the medium could for instance be advantageously coupled to nonlinear detection, in both backward and forward directions. Furthermore, recent applications of wavefront shaping for optical coherence tomography have demonstrated enhanced penetration depth~\cite{Jang2013}, with the latter being conveniently coupled with NLM~\cite{Beaurepaire1999,Bredfeldt2005,Kamali2014} thus showing that addressing linear feedback can be considerably more advantageous.

At last, the expected large spectral width of biological media at mm-depth opens interesting prospective for multi-color nonlinear microscopy, such as  multi-label fluorescence or coherent Raman (CR) imaging. In CR microscopy if the vibrational resonance lies within the spectral width of the medium, only a single wavelength would be necessary to characterize the TM of the system. Such situation may be found at $0.1-0.2~l_t$ depths at which CR microscopy has not yet been able to image~\cite{Wright2007,Ntziachristos2010}. Our observations also suggests that for higher order ($>$2) processes, e.g. three-photon fluorescence, third-harmonic generation or coherent anti-Stokes Raman scattering, enhancements above three orders of magnitude could be generated by using cheaper low resolution SLMs. Because in the present method we are addressing the linearly scattered photons, our conclusions are valid for any nonlinear contrast imaging modality. These results finally pave the way for studies of label-free in-depth imaging in biological media.

\section{Methods}
\subsection{Optical set-up}
Short pulses (130~fs, 800~nm, 76~MHz repetition rate, Mira, Coherent) are steered onto a 256x256 pixels reflective SLM (Boulder Nonlinear Systems). The SLM is imaged on the back focal plane of the focusing lens (0.32~NA, achromatic lens, Thorlabs). The scattering medium is placed between the focusing lens and its focus. The nanoKTP crystals (150~nm diameter) are deposited on a coverslip (170~$\mu$m), and imaged by an objective (40x, 0.75~NA, Nikon) on a 12-bit CMOS camera (Flea3, Point Grey), and on either an electron multiplying charge coupled device (emCCD, QuantEM, Roper Scientific) for the nonlinear feedback scheme, or on a large area photon counting photomultiplier tube (PMT) for the linear feedback scheme (MP 953, PerkinElmer). The SHG signal is spectrally separated with suitable dichroic mirror (560~nm longpass, AHF Analysentechnik), shortpass (700~nm, FESH0700, Thorlabs) and bandpass (400~$\pm$~10~nm, Chroma Technology) filters.

For both linear and nonlinear feedback based TM acquisition, the wavefront phase of a Hadamard base for the incident field is shifted in respect to a reference field (the region in the periphery of the pattern shown in the SLM) in the range [$0-2\pi$] and the nonlinear/linear intensity is recorded by the emCCD/CMOS cameras, depending on the feedback scheme. After measuring all the Hadamard basis, a Fourier transform is applied on the scan of a single basis thus retrieving the phase of the $n$th-basis with respect to the reference field. Once all the Hadamard basis are measured, a unitary transformation is applied to obtain the transmission matrix in the canonical basis~\cite{Popoff2010,Popoff2011a}.

Two different schemes are used for inspection of the signal enhancements depending on the nature of the feedback used for acquiring the TM. In the nonlinear feedback scheme, the wavefront necessary to enhance at a specified position is displayed on the SLM and the nonlinear image taken with the emCCD. In the linear feedback scheme, the retrieved TM contains the information necessary to spatially refocus in the ROI acquired. We then raster-scan the focus using the different elements of the TM at each CMOS pixel within the ROI containing the nanoKTP. In parallel, we collect the SHG signal integrated within the imaged plane by the PMT. A background subtraction was performed on the SHG images.

\subsection{Samples}
Two types of scattering media were used: a commercial diffuser ($10^\circ$ Light Shaping Diffuser, Newport) and a thin TiO$_2$ film (multiply scattering medium). The TiO$_2$ film was fabricated by drop cast from a colloidal solution of amorphous 500~nm-diameter TiO$_2$ plain particles dispersed in water (Corpuscular Inc.). The obtained thickness ($\approx$ 20 $\mu$m) is expected to be about a few $l_t$'s~\cite{Aulbach2011,Andreoli2015}. The collagen fibers were extracted from rat tail tendons as in~\cite{Ait-Belkacem2010} and placed between two coverlips separated by a 120~$\mu$m -thick spacer and filled with agarose solution.


\section*{Acknowledgments}
We thank Esben Andressen and Herve Rigneault for initial tests in their setup and fruitful discussions, Ori Katz for technical assistance, Thierry Gacoin and Ludovic Meyer for providing the nanoKTP particles, and Yannick Foucault for characterizing the TiO$_2$ films. H.B.A. thanks the FEMTO network of Mission pour l'Interdisciplinarit\'{e} (CNRS, France) for its financial support. This works has been supported by Agence Nationale de la Recherche through contracts ANR-10-INBS-04-01 (France-BioImaging infrastructure network), ANR-11-INSB-0006 (France Life Imaging infrastructure network), ANR-15-CE19-0018-01 (MyDeepCARS), by the A*MIDEX project (ANR-11-IDEX-0001-02), and by the European Research Council (grant no. 278025).




\end{document}